\documentclass[12pt]{iopart}
\input epsf

\newcommand{\mbf}[1]{\mbox{\boldmath$#1$\unboldmath}}

\begin{document}

\title{Quantifying spatially heterogeneous dynamics in computer
simulations of glass-forming liquids}

\author{Sharon C. Glotzer\footnote[1]{To whom correspondence should be
addressed. Email: sharon.glotzer@nist.gov} and Claudio
Donati\footnote[2]{Present address: University of Rome ``La
Sapienza'', Rome, Italy.}}

\address{Polymers Division and Center for Theoretical and
Computational Materials Science, National Institute of Standards and
Technology, Gaithersburg, Maryland, USA 20899}

\begin{abstract}
We examine the phenomenon of dynamical heterogeneity in computer
simulations of an equilibrium, glass-forming liquid. We describe
several approaches to quantify the spatial correlation of
single-particle motion, and show that spatial correlations of particle
displacements become increasingly long-range as the temperature
decreases toward the mode coupling critical temperature.
\end{abstract}

\pacs{02.70.Ns, 61.20.Lc, 61.43.Fs}

\submitted{{\noindent \it Proceedings of the 1998 Pisa Workshop on Supercooled
Liquids and Glassy Materials.}}

\section{Introduction}

Liquids cooled toward their glass transition exhibit remarkable
dynamical behavior \cite{ean}.  The initial slowing down of many
liquids at temperatures well above their glass transition temperature
$T_g$ can be described to great extent by the mode coupling theory
\cite{gotze,cummins,mctbook}, which predicts diverging relaxation
times at a dynamical critical temperature $T_c$ despite the absence of
a diverging or even growing static correlation length
\cite{vanblaad}. At the same time, experiments in the temperature
range $T_g < T < T_c$, and simulations in the range $T>T_c$, have
shown that it is possible to select subsets of particles in the liquid
(or monomers in the case of polymers) that move differently from the
bulk on time scales less than the structural relaxation time
\cite{ss91,rev_vigo97, cicerone, sillescu}. The question then arises as to
whether glass-forming liquids exhibit {\it spatially heterogeneous}
dynamics, and if so, is there a growing length scale associated with
this dynamical heterogeneity?

We have employed three complementary computational approaches to
address this question:
\begin{enumerate}
\item {\it Subset approach.} In this approach
\cite{kdppg,ddkppg,dgpkp}, we monitor the displacement of each
particle in a time window $\Delta t$, rank the displacements from
largest to smallest, and examine spatial correlations within subsets
of particles exhibiting either extremely large or extremely small
displacements. In this way, we have demonstrated that particles of
similar mobility form clusters that grow with decreasing temperature
$T$. Moreover, particles exhibiting the largest displacements within a
time window in the late-$\beta$/early-$\alpha$ relaxation regime
(``mobile'' particles) form low-dimensional clusters that grow and
percolate at the mode coupling transition temperature $T_c$.

\item {\it Displacement-displacement correlation function approach.}
In this approach \cite{gdp,pdg,dgp}, we monitor the displacement of
each particle in a time window $\Delta t$ as above, and calculate a
bulk, equilibrium correlation function that measures the correlations between the
local fluctuations in the particle displacements. Because this
approach does not rely on defining subsets via arbitrary thresholds or
cutoffs, it may prove more amenable to future analytical treatment.

\item {\it First passage time approach.}  In this approach
\cite{adg}, we monitor the first passage time $\tau$ for each
particle to move a distance $\epsilon$, and then for a given
$\epsilon$, examine spatial correlations between the different
$\tau$ values either by grouping particles into subsets of similar
$\tau$, or by calculating a $\tau$-$\tau$ correlation function using
all the particles.  This approach shows that particles with similar
first passage times to travel any $\epsilon$ are spatially
correlated. It also allows us to identify particular values of
$\epsilon$ for which the first passage times are most correlated,
and thereby extract detailed information on, e.g., the cage size.

\end{enumerate}

In this paper, we present results from the first two approaches
outlined above, and show that as an equilibrium Lennard-Jones liquid
approaches $T_c$, correlations between local fluctuations in particle
displacements develop, grow, become increasingly long-ranged, and
appear to diverge at $T_c$. Similar results for a simulated polymer
blend are presented elsewhere \cite{bdgb}.

\section{Theory}

Consider a liquid confined to a volume $V$, consisting of $N$
identical particles, each with no internal degrees of freedom.  Let
the position of each particle $i$ be denoted ${\mbf r}_i$.  The motion
of a particle $i$ may be described by its displacement $\mu_i(t,\Delta
t)\equiv |{\mbf r}_i(t+\Delta t) - {\mbf r}_i(t)|$ over some interval of
time $\Delta t$, starting from a reference time $t$.  The question of
whether particle motions in a liquid are correlated can be addressed
by modifying the usual definition of the density-density correlation
function $G({\mbf r})$ (which measures spatial correlations of
fluctuations in local density away from the average value
\cite{hansen,stanley}) so that the contribution of each particle $i$
to $G({\mbf r})$ is weighted by $\mu_i$. Thus, we define a
``displacement-displacement'' correlation function
\cite{gdp,pdg,dgp,spinglass},
\begin{equation}
\label{guu}
G_u({\mbf r},\Delta t)= 
\int d{\mbf r'} \Bigl\langle \bigl [ u({\mbf r'}+{\mbf r},t,\Delta t)  
- \langle u \rangle \bigr ] \, \bigl [ u({\mbf r'},t,\Delta t)  
- \langle u \rangle \bigr ] \Bigl\rangle,
\end{equation}
where
\begin{equation}
u({\mbf r},t,\Delta t)
=\sum_{i=1}^N 
\mu_i(t,\Delta t)
\, \delta\bigl({\mbf r}-{\mbf r}_{i}(t)\bigr).
\label{u}
\end{equation}
and where $\langle \cdots \rangle$ denotes an ensemble average.
$G_u({\mbf r},\Delta t)$ measures the correlations of fluctuations of
local displacements away from their average value.  Since we are
considering an equilibrium liquid, and because $G_u$ is defined as an
ensemble average, $G_u$ does not depend on the choice of the reference
time $t$.  For the same reasons, $\langle u \rangle \equiv \langle
u({\mbf r}, t, \Delta t) \rangle$ only depends on $\Delta t$ for a
homogeneous, equilibrium liquid.

$G_u({\mbf r},\Delta t)$ can be separated into self and distinct
parts, so that we can identify a spatial correlation function
$g_u({\mbf r},\Delta t)$ analogous to the static pair correlation
function $g({\mbf r})$ conventionally used to characterize the
structure of a liquid \cite{hansen}:
\begin{equation}
G_u({\mbf r},\Delta t) = {\bigl \langle N \bigr \rangle} {\overline
{\mu^2}} \, \delta({\mbf r}) + \langle u \rangle \langle U
\rangle \, [g_u({\mbf r},\Delta t) - 1],
\end{equation}
where
\begin{equation}
g_u({\mbf r},\Delta t) =
{{1}\over{\langle u \rangle \langle U \rangle}} \Biggl \langle
\sum_{i=1}^N \sum_{{\scriptstyle j=1} \atop {\scriptstyle j\neq i}}^{N}
\mu_i(t,\Delta t) \, \mu_j(t,\Delta t) \,
\delta\bigl({\mbf r}+{\mbf r}_j(t)-{\mbf r}_i(t)\bigr)
\Biggl \rangle. 
\end{equation}
Here we have defined the ``total displacement'' $U$ as $U(t,\Delta
t)=\int d{\mbf r} \, u({\mbf r},t,\Delta t)$, and its ensemble average
$\langle U \rangle \equiv \langle U(t,\Delta t)\rangle$. We have also
defined ${\bar \mu} \equiv \bigl \langle \frac {1}{N} \sum_{i=1}^N
\mu_i(t,\Delta t) \bigr \rangle$, and ${\overline {\mu^2}} \equiv
\bigl \langle \frac {1}{N} \sum_{i=1}^N \mu_i^2(t,\Delta t) \bigr
\rangle$.  In a constant-$N$ ensemble, $\langle u \rangle$ and
$\langle U \rangle$ are readily evaluated as $\bigl \langle u \bigr
\rangle = \rho \, {\bar \mu}$ and $\bigl \langle U \bigr \rangle = N
\, {\bar \mu}$.  Note that $g_u(r,\Delta t)$ is different from the
distinct part of the time-dependent van Hove correlation function
$g(r,t)$, because it correlates information between pairs of particles
using information about the position of each particle at {\it two}
different times. Thus $g_u(r,\Delta t)$ is in some sense a four-point
correlation function, whereas $g(r,t)$ is a two-point correlation
function.

A ``structure factor'' for the particle displacements can be defined
as
\begin{eqnarray}
S_u({\mbf q},\Delta t) \equiv
\Biggl \langle \frac{1}{N {\overline {\mu^2}} } \sum_{i=1}^N
\sum_{j=1}^N \mu_i(t,\Delta t) \mu_j(t,\Delta t) \exp \left[-i{\mbf q}\cdot
({\mbf r}_i(t) - {\mbf r}_j(t))\right] \Biggr \rangle.
\end{eqnarray}

The fluctuations of $U$ are
related to the volume integral of $G_u({\mbf r},\Delta t)$:
\begin{equation}
\Bigl \langle \bigl [ U - \bigl \langle U
\bigr \rangle \bigr ]^2 \Bigr \rangle = \int d{\mbf r} \, G_u({\mbf
r},\Delta t) \equiv \bigl \langle U \bigr \rangle \bigl
\langle u \bigr \rangle kT \kappa_u,
\label{flucuu}
\end{equation}
where $k$ is Boltzmann's constant.  Recall that the volume integral of
the density-density correlation function $G({\mbf r})$ is proportional
to the isothermal compressibility $\kappa$, which diverges at a
conventional critical point because $G({\mbf r})$ becomes
long-ranged. Likewise, the fluctuations in $U$ will provide
information regarding the range of $G_u({\mbf r},\Delta t)$.  Thus, in
analogy with conventional critical phenomena, we have defined an
isothermal ``susceptibility'' $\kappa_u$.

For a fixed choice of $\Delta t$, note that if the displacement were
always the same for every particle, then $g_u({\mbf r},\Delta t)$ and
$g({\mbf r})$ would be identical for all ${\mbf r}$ (and $S_u({\mbf
q},\Delta t)$ and $S({\mbf q})$ would be identical for all ${\mbf
q}$).  Hence, it is deviations of $g_u({\mbf r},\Delta t)$ ($S_u({\mbf
q},\Delta t)$) from $g({\mbf r})$ ($S({\mbf q})$) that will inform us
of correlations of fluctuations of local displacements away from the
average value, that are in excess of those that would be expected
based on a knowledge of $g({\mbf r})$ or $S({\mbf q})$ alone.

\section{Simulation Details}

We have measured the spatial correlations in particle displacements
using data obtained \cite{kdppg,ddkppg,dgpkp} from a molecular
dynamics simulation of a model Lennard-Jones glass-former.  The system
is a three-dimensional binary mixture (80:20) of 8000 particles
interacting via Lennard-Jones interaction parameters \cite{units}.  We
analyze data from seven $(\rho,P,T)$ state points that lie on a line
in the $P,T$ plane, approaching the mode-coupling dynamical critical
temperature $T_c \approx 0.435$ at a pressure $P_c \approx 3.03$ and
density $\rho_c \approx 1.2$ \cite{kob} (all values are quoted in
reduced units \cite{units}).  The highest and lowest temperature state
points simulated are $(\rho = 1.09, P=0.50, T=0.550)$ and
$(\rho=1.19,P=2.68,T=0.451)$, respectively. Following equilibration at
each state point, the particle displacements are monitored in the NVE
ensemble for up to $1.2 \times 10^4$ Lennard-Jones time units (25.4 ns
in argon units) for the coldest $T$. Complete simulation details may
be found in \cite{dgpkp}.

For all seven state points, a ``plateau'' exists in both the mean
square displacement and the intermediate scattering function
$F_s({\mbf q},t)$ as a function of $t$ \cite{dgpkp}.  The plateau
separates an early time ballistic regime from a late time diffusive
regime, and indicates ``caging'' of the particles typical of low $T$,
high density liquids.  The same model liquid, simulated along a
different path toward the same mode coupling critical point, was found
to be well described by the ideal mode coupling theory \cite{kob}.
For the present (larger) simulation, the $\alpha$-relaxation time
$\tau_{\alpha}$ that describes the long-time decay of $F_s({\mbf
q},t)$ at the value of $q$ corresponding to the first peak in the
static structure factor $S({\mbf q})$ increases by 2.4 orders of
magnitude, and follows a power law $\tau_{\alpha} \sim
(T-T_c)^{-\gamma}$, with $T_c \simeq 0.435$ and $\gamma \simeq
2.8$. The diffusion coefficient $D$ is found to behave as $D \sim
(T-0.435)^{2.13}$. The simulated liquid states analyzed here therefore
exhibit the complex bulk relaxation behavior characteristic of a
supercooled liquid approaching its glass transition.

\section{Results}

Throughout the following subsection, all quantities are calculated
using all 8000 particles in the system. Results do not change
substantially if the minority particles are not included in the
analysis. In the cluster analysis in Subsection {\it
\ref{seccluster}}, only the majority particles are included in the
analysis.  However, our results do not change quantitatively when the
minority particles are included. Recall that in the mode coupling
analysis of this model liquid by Kob and Andersen, the same mode
coupling transition temperature was found for the majority particles
as for the minority particles \cite{kob}.

\subsection{Displacement-displacement correlation function}

\begin{figure}
\hbox to\hsize{\epsfxsize=0.5\hsize\hfil\epsfbox{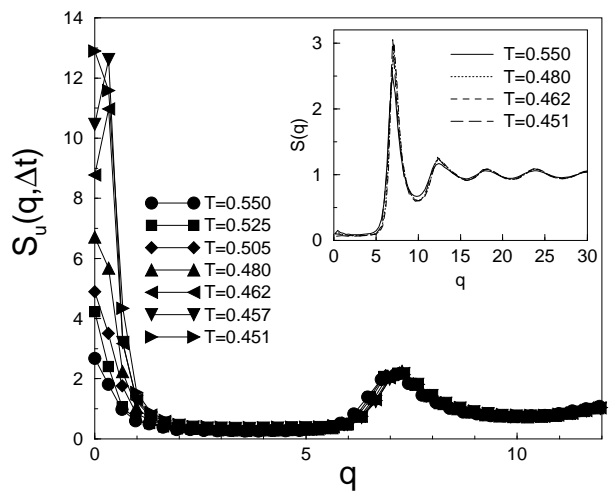}\hfil}
\caption{Displacement ``structure factor'' $S_u(q,\Delta t)$ for
different $T$. The $q=0$ values are obtained from $S_u(q=0,\Delta t) =
({\overline \mu}^2/{\overline {\mu^2}}) \,
\rho kT\kappa_u$.  INSET: Static structure
factor $S(q)$ for four different $T$.}
\label{figsuu}
\end{figure}

$S_u( q,\Delta t)$ is shown for different $T$ in Fig.~\ref{figsuu},
calculated for a value of $\Delta t$ chosen on the order of
$\tau_{\alpha}$ for each $T$. For intermediate and large $q$,
$S_u(q,\Delta t)$ coincides with the static structure factor $S({\mbf
q}) \equiv \Bigl \langle \frac{1}{N} \sum_{i=1}^N \sum_{j=1}^N \exp
\left[{-i{\mbf q}\cdot ({\mbf r}_i - {\mbf r}_j)} \right]\Bigr
\rangle$ (see inset).  However, for $q\rightarrow 0$ a peak in
$S_u(q,\Delta t)$ develops and grows with decreasing $T$, suggesting
the presence of increasing long range correlations in $G_u({\mbf
r},\Delta t)$. No such growing peak at $q=0$ appears in the static
structure factor $S(q)$ (cf. inset), indicating the absence of long
range correlations in the particle positions.
Indeed, a comparison of $g_u(r,\Delta t)$ and $g(r)$ shows that at
every $T$ simulated, the particle displacements are more correlated
than their positions. At the lowest $T$, this excess correlation
persists for values of $r$ up to 6 interparticle distances \cite{dgp}.

\begin{figure}
\hbox to\hsize{\epsfxsize=0.5\hsize\hfil\epsfbox{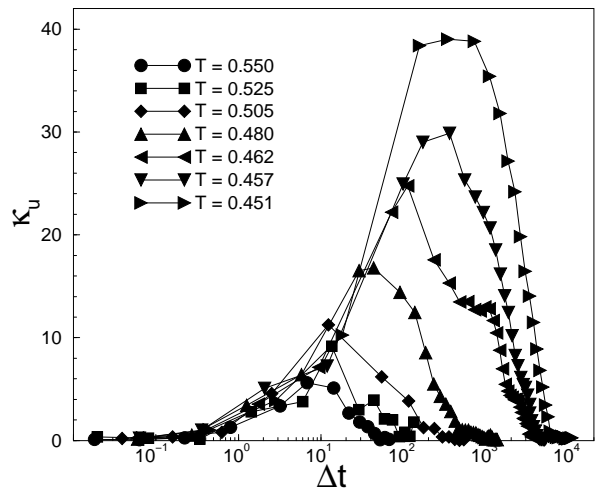}\hfil}
\caption{$\kappa_u(\Delta t)$ as a function of $\Delta t$ for different temperatures.}
\label{figchidt}
\end{figure}

In Fig.\ref{figsuu} $\Delta t$ is chosen on the timescale of the
$\alpha$-relaxation time. The significance of the time interval
$\Delta t$ can be seen by examining the ``susceptibility'' $\kappa_u
(\Delta t)$.  Here we have calculated $\kappa_u(\Delta t)$ directly
from the fluctuations in the total displacement $U$.
Fig.~\ref{figchidt} demonstrates four important points: (i) As $\Delta
t \to 0$, $\kappa_u \to 0$. Note that in the limit $\Delta t \to 0$,
$\mu_i(t,\Delta t)/\Delta t$ is equal to the magnitude of the
instantaneous velocity.  Thus correlations cannot be observed by
looking at a ``snapshot'' of the system, that is, by measuring
correlations in the instantaneous velocity \cite{hiwatari}.  (ii)
Initially, as $\Delta t$ increases, $\kappa_u$ increases. Thus the
spatial correlation of particle displacements develops over time.
(iii) There is a time window $\Delta t^*$ during which $\kappa_u$ is
maximum. Both the maximum value of $\kappa_u$ and corresponding time
window $\Delta t^*$ increase with decreasing $T$. (iv) As $\Delta t
\to \infty$, $\kappa_u$ decreases.  Thus in the diffusive regime,
where the particles act like Brownian particles, there are no
correlations in the particle motions.

\begin{figure}
\hbox to\hsize{\epsfxsize=0.5\hsize\hfil\epsfbox{
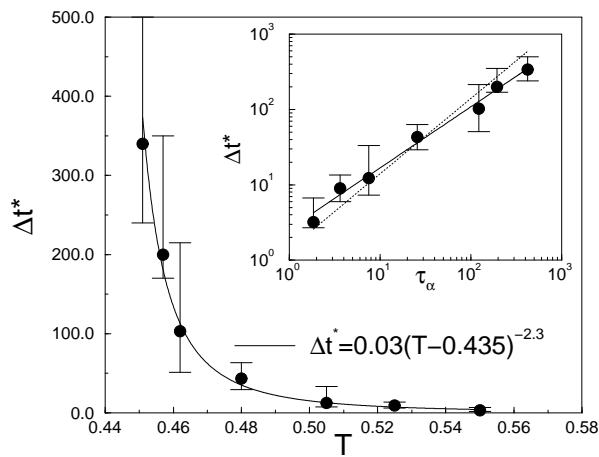}\hfil}
\caption{Time when the fluctuations in the particle displacements are
largest, plotted vs. temperature.  The solid curve is a fit to $\Delta
t^* \sim (T-0.435)^{-2.3}$. INSET: $\Delta t^*$ vs. $\tau_{\alpha}$,
plotted log-log. The solid line has slope $0.81$, and represents the
best fit curve through the data. The dashed line has slope $1.0$, and
is not nearly as good a fit.  That $\Delta t^*$ does not scale
linearly with $\tau_{\alpha}$ is further supported by the fact that an
exponent less than $2.6$ describing the apparent divergence of
$\tau_{\alpha}$ with $T-T_c$ can be excluded.}
\label{figdt}
\end{figure}

Fig.~\ref{figdt} shows that the time $\Delta t^*$ at which the
fluctuations in the particle displacements are a maximum follows a
power law with T: an excellent fit to the form $\Delta t^* \sim
(T-T_c)^{-\gamma}$ is obtained when $T_c=0.435 \pm 0.005$, and yields
$\gamma = 2.3 \pm 0.3$, where the highest value of $\gamma$ is
obtained with the lower bound on $T_c$.  Our estimated value for
$\gamma$ differs from that describing the divergence at $T_c$ of the
structural relaxation time $\tau_{\alpha}$ ($\gamma \simeq 2.8$), but
(within our numerical uncertainty) cannot be distinguished from the
exponent governing the apparent vanishing of the diffusion coefficient
$D$ ($\gamma \simeq 2.1 \pm 0.1$).

\begin{figure}
\hbox to\hsize{\epsfxsize=0.5\hsize\hfil\epsfbox{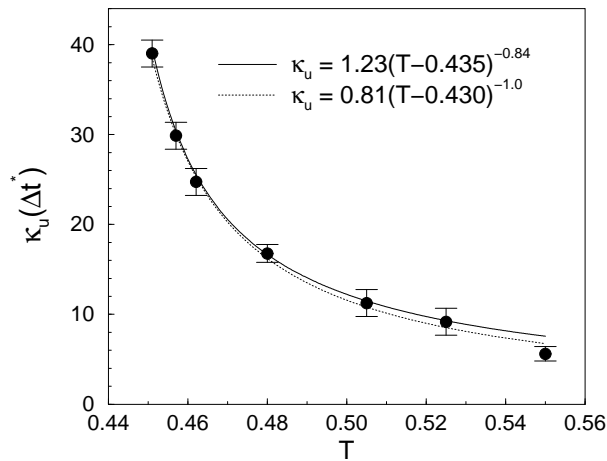}\hfil}
\caption{``Susceptibility'' $\kappa_u(\Delta t^*)$ versus temperature.
The solid and dashed curves are power law fits to the data, as indicated
in the figure.}
\label{figchiT}
\end{figure}

In Fig~\ref{figchiT} we show the $T$-dependence of $\kappa_u(\Delta
t^*)$. We find that $\kappa_u(\Delta t^*)$ grows monotonically with
decreasing $T$, indicating that the range of the correlation measured
by $g_u(r,\Delta t^*)$ is growing with decreasing $T$. The data can be
fitted extremely well with a power law $\kappa_u(\Delta t^*) \sim
(T-T_c)^{-\gamma}$ with $T_c = 0.435$, and yields $\gamma = 0.84$.  (As
shown in the figure, a larger (closer to Curie-Weiss-type) exponent
can be obtained by fitting the data with a power law that diverges at
a slightly lower value of $T_c$.) Thus the previously determined mode
coupling critical temperature coincides, within our numerical error,
with the temperature where $G_u(r,\Delta t)$ becomes long ranged.

\begin{figure}
\hbox to\hsize{\epsfxsize=0.5\hsize\hfil\epsfbox{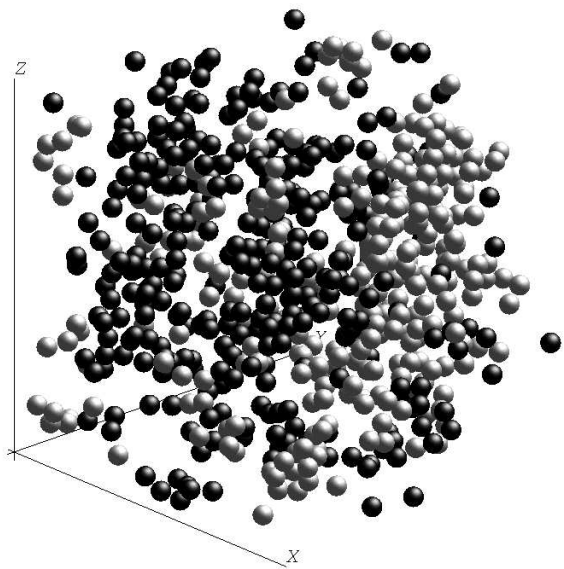}\hfil}
\caption{Particles exhibiting the 5\% largest (light) and 5\% smallest
(dark) displacements in $\Delta t^*$ starting from an arbitrary time
origin, at $T=0.451$. The clustering of particles of similar mobility
is evident in the figure.}
\label{figextreme}
\end{figure}

The displacement-displacement correlation function measures the
tendency for particles of similar mobility to be spatially correlated
(cf. Fig.~\ref{figextreme}).
To extract a correlation length from the data, we have attempted to
fit $S_u(q,\Delta t^*)$ using an Orstein-Zernike form, $S_u(q,\Delta
t^*)\propto 1/(1+\xi^2 q^2)$. This form assumes that $G_u({\mbf
r},\Delta t^*)$ is asymptotically proportional to $e^{-r/\xi}/r$,
where $\xi$ is the correlation length \cite{stanley}. We find that
this form fits well at the highest $T$ but fails to fit the data on
approaching $T_c$.  The data can instead be fitted at all $T$ using
the more general form $S_u(q,\Delta t^*)\propto 1/(1+(\xi q)^\eta)$.
The best fit gives $\eta=1.99$ and 3.54 for our highest and lowest
$T$, respectively. However, because $\eta$ depends on $T$, $\xi$ does
not show a definite trend with temperature, giving values of roughly
three particle diameters.  Larger simulations may be required to
accurately determine the correct functional form for $S_u$ at small
$q$.  Nevertheless, the data show unambiguously that as $T \to T_c$,
spatial correlations between the displacements of particles arise and
grow and become long-ranged at $T_c$. To obtain further evidence for a
growing dynamical length scale, we turn now to the subset method
described in the Introduction \cite{length}.

\subsection{Cluster analysis of mobile particles}
\label{seccluster}

Examination of the distribution of particle displacements (the self
part of the usual time-dependent van Hove correlation function
\cite{hansen}) immediately shows the problems inherent in defining
subsets of particles of extremely low mobility. Over the time range
coincident with the plateau in the mean square displacement, this
function exhibits a narrow peak containing most of the particles,
and a long tail to large distances containing a small percentage of
the particles. The non-Gaussian parameter $\alpha_2(t)$ has been used
to identify a time when this tail is most pronounced \cite{kdppg}. At
that time, approximately 95\% of the particles are contained in the
peak, and about 5\% are contained in the tail, regardless of
temperature. Because of this tail, the most mobile particles in the
liquid ``distinguish'' themselves from the bulk, while it is more
difficult to define a threshold with which to identify particles
exhibiting extremely small displacements (``immobile'' particles).

\begin{figure}
\hbox
to\hsize{\epsfxsize=0.5\hsize\hfil\epsfbox{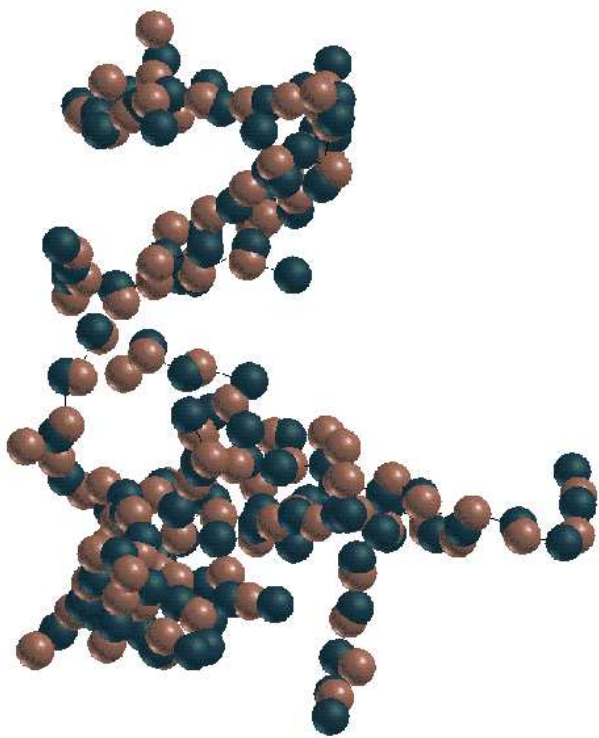}\hfil}
\caption{A large cluster of mobile particles found at $T=0.451$, at
two different times. Pink denotes the particles at $t=0$ and blue
denotes the particles at time $\Delta t$ later, where $\Delta t$ is
slightly less than $\Delta t^*$. The small black lines connect the
same particle at the two times. Thus between $t=0$ and $\Delta t$,
each particle shown has moved almost one interparticle distance.  This
cluster can be decomposed into a number of elementary ``cooperatively
rearranging strings'', defined as minimal groups of particles that
follow one another \cite{ddkppg,addg}.}
\label{figdino}
\end{figure}

By examining the {\it maximum} displacement achieved by a particle in a time
window $\Delta t$, a somewhat better definition of immobility can be
achieved.  It was shown in Ref.~\cite{dgpkp} that particles exhibiting
the smallest displacement defined in this way are spatially
correlated, and form relatively compact clusters. Using this
definition of maximum displacement, which for the most mobile
particles gives the same results as the simple definition of
displacement used above, we find that highly mobile particles also
cluster. These clusters are very ramified, and are composed of
smaller ``strings'' of particles that follow one another
(cf. Fig.~\ref{figdino}).

\begin{figure}
\hbox to\hsize{\epsfxsize=0.5\hsize\hfil\epsfbox{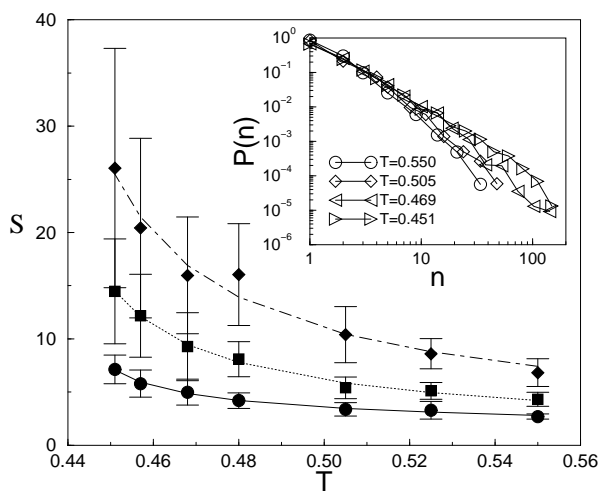}\hfil}
\caption{Mean cluster size $S$ plotted versus $T$, for subsets
containing 3\% (circles), 5\% (squares), and 7\% (diamonds) of the
most mobile particles at $T=0.451$. The lines are power law fits $S \sim
(T-T_p)^{-\gamma}$.  Best fit parameters are $T_p = 0.440, 0.431$ and
$0.428$, respectively, and $\gamma = 0.397, 0.687$, and $0.741$,
respectively. INSET: Distribution of the size $n$ of clusters of
mobile particles for four different $T$.}
\label{figperc}
\end{figure}

In this subsection, we are interested in the growth of clusters of
highly mobile particles with decreasing temperature.  To this end, we
have constructed clusters of nearest neighbor particles whose
maximum displacement in a time window $\Delta t \sim \Delta t^*$ falls within
the top 3\%, 5\% and 7\% of all displacements \cite{dgpkp}. The
distribution of clusters of size $n$ constructed using the top 5\% of
the particles is shown in the inset of Fig.~\ref{figperc} for four
different $T$.  Also shown in Fig.~\ref{figperc} is the $T$-dependence
of the mean cluster size $S = \sum n^2 P(n) / \sum nP(n)$ for each
subset \cite{stauffer}. We find that for each subset $S \sim
(T-T_p)^{-\gamma}$ with $T_p = 0.440, 0.431$, and $0.428$, all close
to $T_c$. (Note that as the number of
particles contained in the subset increases, $T_p$ {\it decreases}.)
For the subset containing 5\%, the data fall on a straight line when
plotted log-log against $T-T_c$.  Thus, we conclude that $T_c$ appears
to coincide with a percolation transition of the most mobile particles
in the liquid.

The presence of clusters of mobile particles whose size grows with
decreasing temperature contributes to the growing range of the
displacement-displacement correlation function.  Thus we can use the
mean cluster size to give a rough estimate of the length scale on
which particle motions are correlated. In our coldest simulation, the
mean cluster size is approximately $10$ particles. Since the clusters have a
fractal dimension of approximately 1.75 \cite{dgpkp}, that implies a
mean radius of gyration larger than 3 particle diameters.  The largest
cluster observed at $T=0.451$ contains over 100 particles, and thus
has a radius of gyration that exceeds the size of our simulation box
at that $T$.

It is important to note that while the mean size of clusters of mobile
particles appears to diverge at $T_c$, the smaller, cooperatively
rearranging ``strings'' that make up these clusters also grow with
decreasing $T$, {\it but are still finite at $T_c$} \cite{ddkppg}. In the
temperature regime studied here, the distribution of string lengths is
exponential, and the mean string length appears to diverge at a
temperature substantially lower than $T_c$.  Thus while the mode
coupling transition may coincide with a percolation transition of
clusters of highly mobile particles, our data is compatible with the
idea of Adam and Gibbs that the ideal glass transition at $T_o < T_c$
is associated with the growth and possible divergence of the minimum
size of a cooperatively rearranging region (``strings'' in the present
simulation).  This idea will be further explored elsewhere
\cite{addg}.

\section{Discussion}

In this paper, we have defined a bulk correlation function that
quantifies the spatial correlation of single-particle displacements in
a liquid, and we have examined spatial correlations within subsets of
highly mobile particles. While the first approach has the advantage of
allowing a direct calculation of spatial correlations in local
particle motions without having to define arbitrary subsets, the
second approach has the advantage of permitting a geometrical analysis
of clusters of highly mobile and highly immobile particles.  Using
both approaches, we have shown in computer simulations of an
equilibrium Lennard-Jones liquid that the displacements of particles
are spatially correlated \cite{othercomp} over a range and time scale
that both grow with decreasing $T$ as the mode coupling transition is
approached.  While mode coupling theory currently makes no predictions
concerning a growing dynamical correlation length, calculation of the
vector analog of the displacement-displacement correlation function
should be tractable within the mode-coupling framework. Finally, the
displacement-displacement correlation function has allowed us to
identify a fluctuating dynamical variable U whose fluctuations become
longer ranged and appear to diverge at $T_c$. In this regard, $U$ is
behaving much like a static order parameter on approaching a
second-order static critical point, suggesting the possibility that we
can obtain insights into the nature of glass-forming liquids using an
extension to dynamically-defined quantities of the framework of
ordinary critical phenomena.


\begin{thebibliography}{999}
\bibitem{units} The Lennard-Jones interaction parameters
$\epsilon_{\alpha\beta}$ and $\sigma_{\alpha\beta}$ are given by:
$\epsilon_{AA}=1.0$, $\epsilon_{AB}=1.5$, $\epsilon_{BB}=0.5$,
$\sigma_{AA}=1.0$, $\sigma_{AB}=0.8$, $\sigma_{BB}=0.88$. Lengths are
defined in units of $\sigma_{AA}$, temperature $T$ in units of
$\epsilon_{AA}/k_B$, and time $t$ in units of
$\sqrt{\sigma_{AA}^2m/\epsilon_{AA}}$. Both types of particles are
taken to have the same mass.
\end{thebibliography}
\end{document}